\documentclass[12pt]{article}
\usepackage{axodraw,bbold}

\catcode`@=12
\begin{document}
\thispagestyle{empty}
\begin{flushright} 
UCRHEP-T439\\ 
September 2007\
\end{flushright}
\vspace{0.5in}
\begin{center}
{\LARGE	\bf Near Tribimaximal Neutrino Mixing\\ with $\Delta(27)$ Symmetry\\}
\vspace{1.5in}
{\bf Ernest Ma\\}
\vspace{0.2in}
{\sl Physics Department, University of California, Riverside, 
California 92521 \\}
\vspace{1.5in}
\end{center}

\begin{abstract}\
The discrete subgroup $\Delta(27)$ of $SU(3)$ has the interesting 
multiplication rule $3 \times 3 = \bar{3} + \bar{3} + \bar{3}$, which 
is used to obtain near tribimaximal neutrino mixing. Using present 
neutrino oscillation data as input, this model predicts that the 
effective mass $m_{ee}$ measured in neutrinoless double beta 
decay will be 0.14 eV.
\end{abstract}

\newpage
\baselineskip 24pt

The non-Abelian discrete subgroup $\Delta(12)$ of $SU(3)$ \cite{su3}, more 
familiarly known as $A_4$ \cite{mr01,bmv03}, has been shown to be useful 
\cite{m04,af05,bh05,m07-1,m07-2} for obtaining the tribimaximal mixing 
\cite{hps02} of neutrinos, in good agreement \cite{gm07} with 
present data.  In the basis where the charged-lepton mass matrix 
${\cal M}_l$ is diagonal, the Majorana neutrino mass matrix is given by
\cite{m04}
\begin{equation}
{\cal M}_\nu = \pmatrix{x+y+z & -x & -x \cr -x & y & z \cr -x & z & y}
\end{equation}
which is form diagonal, i.e. it is diagonalized by
\begin{equation}
U_3 = \pmatrix{\sqrt{2/3} & \sqrt{1/3} & 0 \cr -\sqrt{1/6} 
& \sqrt{1/3} & -\sqrt{1/2} \cr -\sqrt{1/6} & \sqrt{1/3} & \sqrt{1/2}},
\end{equation}
independent of its three mass eigenvalues
\begin{equation}
m_1 = 2x + y + z, ~~ m_2 = -x + y + z, ~~ m_3 = y - z.
\end{equation}
Since the values of $x,y,z$ are arbitrary, neutrino masses are not predicted 
in such a scheme.  [However, they can be restricted in two special cases: 
(A) $y=2x$ \cite{af05}, and (B) $(x+y)^2=(2x-y)(x+z)$ \cite{bh05}.]

In this paper, using the discrete group $\Delta(27)$ \cite{bgg84,m06-1,mkr07} 
which is next in the sequence of $\Delta(3n^2)$ subgroups of $SU(3)$, it will 
be shown that ${\cal M}_\nu$ deviates slightly from Eq.~(1) such that the 
change of $\tan^2 \theta_{12}$ from 0.5 to 0.45 allows the prediction of 
$m_{ee} = 0.14$ eV for the effective neutrino mass in neutrinoless double 
beta decay, a value accessible in the next generation of such experiments 
\cite{aee07}.

The non-Abelian discrete group $\Delta(27)$ has 27 elements divided into 11 
equivalence classes.  It has 9 one-dimensional irreducible representations 
${\bf 1_i} (i=1,...,9$) and 2 three-dimensional ones {\bf 3} and 
$\bar{\bf 3}$.  Its character table and the 27 defining $3 \times 3$ matrices 
are given in Ref.~\cite{m06-1}. Its group multiplication rules are
\begin{equation}
{\bf 3} \times {\bf 3} = \bar{\bf 3} + \bar{\bf 3} + \bar{\bf 3},
~~{\rm and}~~{\bf 3} \times \bar{\bf 3} = \sum^9_{i=1} {\bf 1_i},
\end{equation}
where
\begin{eqnarray}
&& {\bf 1_1} = 1 \bar 1 + 2 \bar 2 + 3 \bar 3, ~~~{\bf 1_2} = 1 \bar 1 + 
\omega 2 \bar 2 + \omega^2 3 \bar 3, ~~~{\bf 1_3} = 1 \bar 1 + \omega^2 2 
\bar 2 + \omega 3 \bar 3, \\
&& {\bf 1_4} = 1 \bar 2 + 2 \bar 3 + 3 \bar 1, ~~~{\bf 1_5} = 1 \bar 2 + 
\omega 2 \bar 3 + \omega^2 3 \bar 1, ~~~{\bf 1_6} = 1 \bar 2 + \omega^2 2 
\bar 3 + \omega 3 \bar 1, \\
&& {\bf 1_7} = 2 \bar 1 + 3 \bar 2 + 1 \bar 3, ~~~{\bf 1_8} = 2 \bar 1 + 
\omega^2 3 \bar 2 + \omega 1 \bar 3, ~~~{\bf 1_9} = 2 \bar 1 + \omega 3 
\bar 2 + \omega^2 1 \bar 3,
\end{eqnarray}
with $\omega = \exp(2 \pi i/3)$, i.e. $1+\omega+\omega^2=0$.

Let the lepton doublets $(\nu_i,l_i)$ as well as singlets $l^c_i$ transform 
as ${\bf 3}$ under $\Delta(27)$, then with three Higgs doublets $(\phi^0_i,
\phi^-_i)$ also transforming as ${\bf 3}$, the charged-lepton mass matrix is 
of the form
\begin{equation}
{\cal M}_l = \pmatrix{h_1 v_1 & h_2 v_3 & h_3 v_2 \cr h_3 v_3 & h_1 v_2 & 
h_2 v_1 \cr h_2 v_2 & h_3 v_1 & h_1 v_3}.
\end{equation}
As shown in Ref.~\cite{m06-2}, if $v_1=v_2=v_3$, this is also form 
diagonal, i.e.
\begin{equation}
{\cal M}_l = U_L \pmatrix{(h_1 + h_2 + h_3)v & 0 & 0 \cr 0 & (h_1 + h_2 \omega 
+ h_3 \omega^2)v & 0 \cr 0 & 0 & (h_1 + h_2 \omega^2 + h_3 \omega)v} 
U_R^\dagger,
\end{equation}
where $U_L=U_R$ is the familiar 
\begin{equation}
U_\omega = {1 \over \sqrt{3}}\pmatrix{1 & 1 & 1 \cr 1 & \omega & \omega^2 
\cr 1 & \omega^2 & \omega},
\end{equation}
first introduced by Cabibbo \cite{c78} and Wolfenstein \cite{w78}.

At the same time, with three Higgs triplets $(\xi^{++},\xi^+,\xi^0)$ 
transforming as ${\bf 3}$, the Majorana neutrino mass matrix is of the 
same form as Eq.~(8) but it has to be symmetric, i.e.
\begin{equation}
{\cal M}_\nu = \pmatrix{f_1 u_1 & f_2 u_3 & f_2 u_2 \cr f_2 u_3 & f_1 u_2 & 
f_2 u_1 \cr f_2 u_2 & f_2 u_1 & f_1 u_3}.
\end{equation}
In this basis, the condition for tribimaximal mixing is $u_2=u_3=0$, but 
then the mass eigenvalues become $m_1=f_1 u_1$, $m_2=f_2 u_1$, and 
$m_3=-f_2 u_1$, which are of course not realistic. 
On the other hand, this represents a symmetry limit, and small deviations 
from it will allow the masses to be different, correlated with  
changes in the mixing angles from those of tribimaximal mixing.
In the following, it is shown how present data will predict 
$m_{ee} = 0.14$ eV in the context of this model.

Given the form of Eq.~(11), let it be rewritten as
\begin{equation}
{\cal M}_\nu = \pmatrix{\lambda d & f & e \cr f & \lambda e & d \cr e & d & 
\lambda f} = U_2 \pmatrix{d + \lambda(e+f)/2 & (e+f)/\sqrt{2} 
& \lambda(-e+f)/2 \cr (e+f)/\sqrt{2} & \lambda d & (e-f)/\sqrt{2} \cr 
\lambda(-e+f)/2 & (e-f)/\sqrt{2} & d - \lambda(e+f)/2} U_2^T,
\end{equation}
where
\begin{equation}
U_2 = \pmatrix{1 & 0 & 0 \cr 0 & 1/\sqrt{2} & -1/\sqrt{2} \cr 0 & 1/\sqrt{2} 
& 1/\sqrt{2}} \pmatrix{0 & 1 & 0 \cr 1 & 0 & 0 \cr 0 & 0 & i} = \pmatrix{
0 & 1 & 0 \cr 1/\sqrt{2} & 0 & -i/\sqrt{2} \cr 1/\sqrt{2} & 0 & i/\sqrt{2}},
\end{equation}
then $U_3$ of Eq.~(2) is obtained from $U_\omega$ of Eq.~(10) and the above, 
i.e. $U_3 = U_\omega^\dagger U_2$.
This means that tribimaximal mixing is approximately obtained provided that 
$e,f << d$.

To obtain $\Delta m^2_{sol} << \Delta m^2_{atm}$, set
\begin{equation}
d + {\lambda \over 2} (e+f) = -\lambda d + \delta,
\end{equation}
where $\delta > 0$ is small and $\lambda d > 0$ has been assumed.  Then
\begin{equation}
m_{1,2} = {\delta \over 2} \mp m_0,
\end{equation}
where $m_0 > 0$ with
\begin{equation}
m_0^2 = \left( \lambda d - {\delta \over 2} \right)^2 + {2 \over \lambda^2} 
[(1+\lambda)d - \delta]^2.
\end{equation}
Hence
\begin{equation}
m_2^2-m_1^2 = 2 \delta m_0 > 0,
\end{equation}
and
\begin{equation}
m_3^2 - m_0^2 = {2 d^2 \over \lambda^2} (\lambda^2 -1)(2 \lambda+1).
\end{equation}
The new $\theta_{12}$ is now given by
\begin{equation}
\tan \theta_{12} \simeq {1 \over \sqrt{2}} \left[ 1 + {3 \over 2} \left( 
{1+\lambda \over \lambda^2} \right) \right] \simeq {1 \over \sqrt{2}} \left[ 
1 + {3 \epsilon \over 2} \right],
\end{equation}
where $\lambda = -1 + \epsilon$ has been used.  Using the experimental 
central value of 0.45 for $\tan^2 \theta_{12}$,
\begin{equation}
\epsilon \simeq -0.034
\end{equation}
is obtained.  Since $m_3^2 - m_0^2 \simeq 4 \epsilon d^2$, this means that 
an inverted ordering of neutrino masses is predicted.  Furthermore, since 
$m_{ee} \simeq |d|$, the experimental central value of $2.7 \times 10^{-3}$ 
eV for $|\Delta m^2_{atm}|$ implies
\begin{equation}
m_{ee} = \left| {\Delta m^2_{atm} \over 4 \epsilon} \right|^{1/2} = 
0.14~{\rm eV}.
\end{equation}
If $\tan^2 \theta_{12} = 0.4$ is used instead, then $m_{ee} = 0.1$ eV.

As for $\theta_{13}$, it is given here by
\begin{equation}
\sin \theta_{13} \simeq (\sqrt{2} \sin \theta_{12} - \cos \theta_{12}) 
\left( {e-f \over 2d} \right) \simeq -0.02 \left( {e-f \over d} \right).
\end{equation}
Since only $(e+f)/d \simeq 2\epsilon$ has been determined, there is no 
prediction for $\theta_{13}$ in this model.

In conclusion, the family symmetry $\Delta(27)$ has been discussed in a 
simple model as the origin of the observed mixing pattern of neutrinos.
It is able to describe present data and has a specific prediction of 
the effective neutrino mass, i.e. 0.14 eV, in neutrinoless double beta decay.

This work was supported in 
part by the U.~S.~Department of Energy under Grant No. DE-FG03-94ER40837.

\baselineskip 18pt

\bibliographystyle{unsrt}

\end{document}